\documentstyle[aps,multicol,epsfig]{revtex}

\renewcommand{\narrowtext}{\begin{multicols}{2}
\global\columnwidth20.5pc}
\renewcommand{\widetext}{\end{multicols}
\global\columnwidth42.5pc} \multicolsep = 8pt plus 4pt minus 3pt

\begin{document}

\title{Simple Phase Bias for Superconducting Circuits}
\author{J. B. Majer, J. R. Butcher, J. E. Mooij}
\address{Applied Physics and DIMES, Delft University of Technology\\
Lorentzweg 1, 2628 CJ Delft, The Netherlands}
\date{\today}
\maketitle

\begin{abstract}
A phase-bias tool, based on a trapped fluxoid in a ring, is proposed and demonstrated. It can provide arbitrary 
phase values and is simple to fabricate.
The phase bias has been realized in two superconducting
quantum interference devices, where the critical current versus magnetic flux is shown to be
shifted by a $\pi/2$ and $\pi$. 
\end{abstract}

\narrowtext
Many superconducting devices require an imposed phase difference. Usually this phase bias is applied by means of magnetic flux in a closed superconducting loop. An example is the superconducting quantum interference device (SQUID) \cite{Tinkham}, which needs a phase bias of around $\pi$/2. Several types of superconducting quantum bits rely on a phase difference $\pi$ \cite{Ioffe,Mooij,Blatter}. So-called $\pi$-junctions, where the dependence of the current on the phase difference is shifted by $\pi$, are seen as attractive components in superconducting electronics \cite{Beasley}. $\pi$-junctions have been developed with d-wave superconductors \cite{dwave} and with superconductor-ferromagnet-superconductor junctions \cite{SFS}, but here severe materials problems are encountered. We have developed a surprisingly simple phase-bias tool, based on trapped fluxoids in a closed loop without junctions. It can be fabricated very easily from materials such as niobium or aluminum; any phase shift can be realized. We present a demonstration 
of its performance in SQUID devices with built-in phases of $\pi$ and $\pi$/2.

\begin{figure}[htb]
\begin{center}
\epsfig{figure=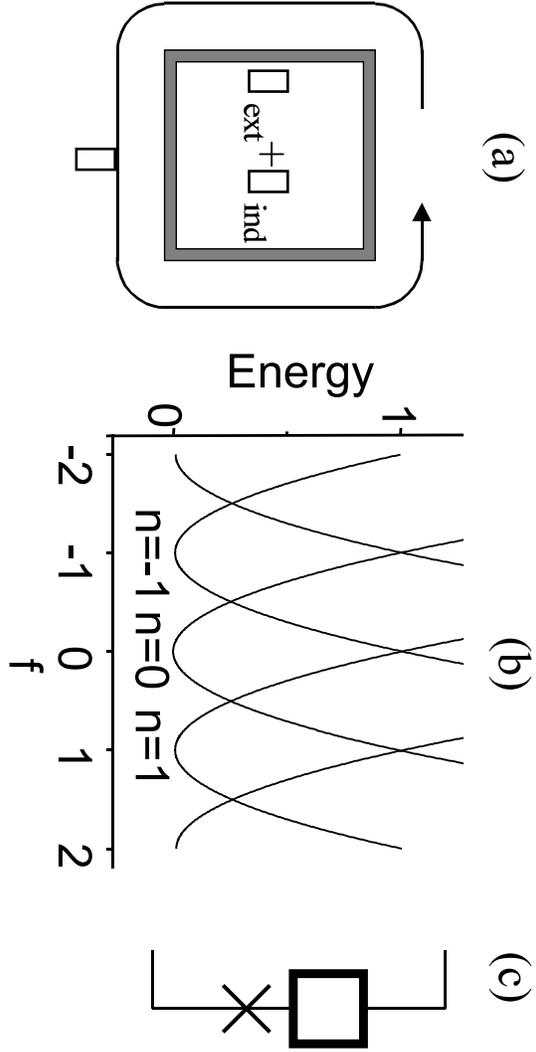, width=14 cm}
\end{center}
\caption{(a) Superconducting ring. $\gamma $ is the phase difference acquired around the ring. The flux in the ring is the externally applied flux plus the self induced flux.
(b) Energy versus externally applied flux $f=\Phi_{\mathrm ext}/\Phi_0$. The energy is normalized to $\Phi_0^2/(L_K(1+\beta))$.  
(c) Circuit element consisting of a ring with one fluxoid and a Josephson junction (cross), which behaves as a $\pi$-junction. }
\label{ring}
\end{figure}

Our approach relies on a mesoscopic superconducting ring [Fig.~\ref{ring}~(a)] with narrow cross section relative to the penetration depth. 
The phase difference along the circumference $\gamma$ is proportional to the current $I$ in the ring. The proportionality factor is given by the kinetic inductance of the ring $L_K$.
\begin{equation}
\gamma=2\pi\frac{L_K}{\Phi_0}I
\end{equation}
where  $\Phi_0$ is the superconducting flux quantum. The kinetic inductance of the ring depends on the London penetration depth $\lambda_L$ and geometrical factors, such as the circumference $s$, width $w$ and height $h$ of the cross section of the wire \cite{orlando}:
\begin{equation}
L_K=\mu_0\lambda_L^2 \frac{s}{w h}
\end{equation}

The flux $\Phi$ in the ring contains two contributions: the externally imposed flux $\Phi_{\mathrm ext}=f\Phi_0$ and the flux $L_GI$ generated by the current, where $L_G$ is the geometrical inductance of the ring. The fluxoid quantization condition \cite{orlando} yields:
\begin{equation}
\gamma=-2\pi\frac{\Phi}{\Phi_0}+2\pi n=2\pi\frac{n-f}{1+\beta}\label{fluxoidcons}
\end{equation}
where $n$ is the integer fluxoid number and $\beta=L_G/L_K$ the ratio between geometrical and kinetic inductance. Changes in $n$ are only possible through phase slip processes, requiring the order parameter to go to zero in a region of the order of the coherence length. The energy required for a phase slip at zero current is approximately equal to $\sqrt{6} I_w\Phi_0/2\pi$ \cite{mccumberhalperin}, where $I_w$ is the critical current of the wire. As critical currents of even narrow wires are typically in the order of 1 mA, the barrier is very high $>10000 K$.

The energy of the the ring has a parabolic dependence on the applied flux $f\Phi_0$ [Fig.~\ref{ring}~(b)]. \begin{equation}
E=\frac{1}{2}L_K I^2+\frac{1}{2}L_GI^2=\frac{\Phi_0^2}{L_K (1+\beta)}(f-n)^2\label{energy}
\end{equation}
If approximately one flux quantum is applied to the ring ($f\approx1$) while cooling down through 
the superconducting phase transition, the ring assumes the lowest energy state $n=1$. 
When the external flux is removed at low temperatures,  the new ground state $n=0$ cannot be reached and the ring remains frozen in the $n=1$ state. 
When a relatively weak superconducting circuit is attached to two contacts on the ring, that circuit experiences a phase shift bias through two channels: the directly picked-up fraction of the total phase and the magnetic flux induced in the loop in the attached circuit. The first contribution is $\gamma a/s$ for a homogeneous ring with circumference $s$ and an enclosed section $a$ between the contacts. It dominates when the kinetic inductance is large, $\beta\ll1$. The contribution from the flux dominates when $\beta\gg1$. In practice both limits can be realized, but intermediate values of $\beta$ are equally useful. For the example of our aluminum SQUID circuits $\beta$ is of order one.
With this ring, new superconducting circuit elements can be made. In Fig.~\ref{ring}~(c) a ring is connected to a standard Josephson junction. This element behaves like a $\pi$-junction in a superconducting network.  

\begin{figure}[htb]
\begin{center}
\epsfig{figure=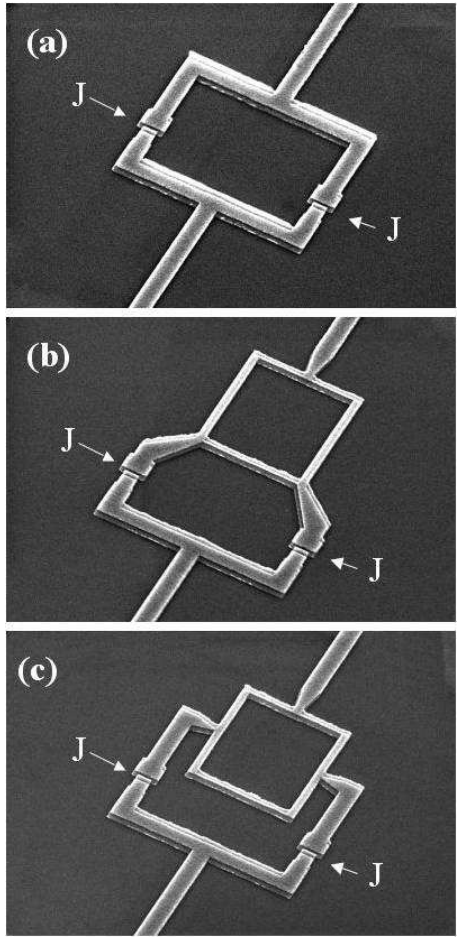, width=6cm, clip=true}
\end{center}
\caption{Scanning electron images of the three devices. (a) shows the reference SQUID, (b) the $\pi$/2-SQUID and (c) the $\pi$-SQUID. The ring has a diameter of 3 $\mu$m. The Josephson junctions are indicated with J.}
\label{SEMpictures}
\end{figure}

This idea is applied to a SQUID. We fabricated three devices [Fig.~\ref{SEMpictures}] on a single substrate 
using standard aluminum shadow evaporation technique. The ring is evaporated in the same layer as the 
Al/Al$_2$O$_3$/Al Josephson junctions, which makes the fabrication very simple. This is illustrated by the fact that the first fabrication attempt was successful. The first device [Fig.~\ref{SEMpictures}~(a)] is a standard SQUID and serves as a reference 
for the two other samples. 
The second device [Fig.~\ref{SEMpictures}~(b)] includes a ring in the SQUID, 
where one quarter of the ring is enclosed. We will refer to this device as the $\pi$/2-SQUID. The third 
device [Fig.~\ref{SEMpictures}~(c)], called the $\pi$-SQUID, encloses half of the ring.       

\begin{figure}[htb]
\begin{center}
\epsfig{figure=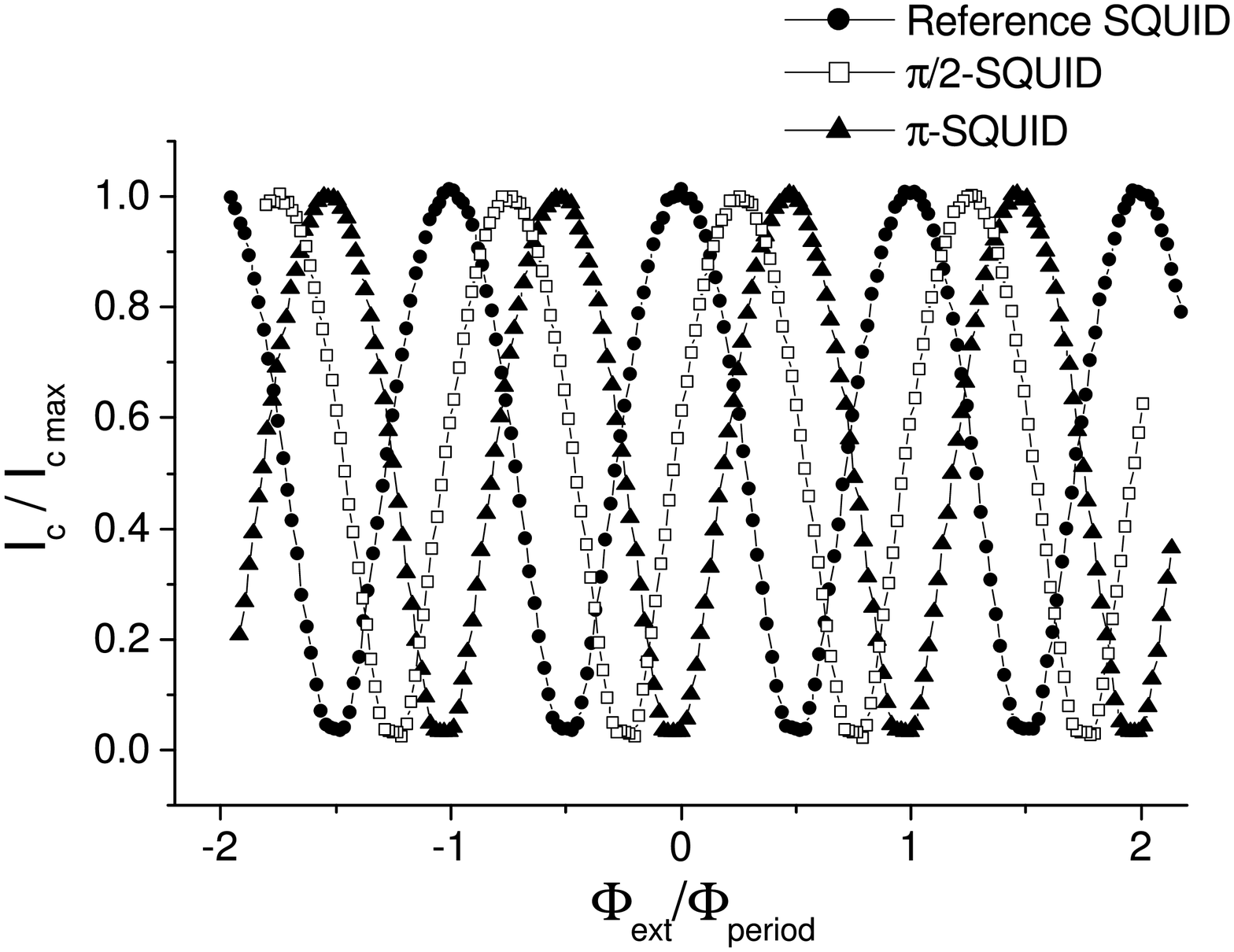, width=11.5 cm, clip=true}
\end{center}
\caption{Normalized critical current as a function of the normalized applied flux. 
The responses of the $\pi/2$ and the $\pi$-SQUID are shifted by approximately a quarter and half a period.}
\label{measurements}
\end{figure}

Measurements were performed in a dilution refrigerator at 60 mK. 
We measured the switching current $I_c$ as a function of the applied magnetic field .
All three devices show regular SQUID oscillations, with maxima and minima clearly shifted with respect to each other. For analysis, the critical current was Fourier analyzed. Results are given in table \ref{fit}. In Fig.~\ref{measurements} the normalized critical current is plotted as a function of the normalized applied flux. Clear shifts of approximately a quarter period for the $\pi/2$-SQUID and a half period for the $\pi$-SQUID are observed.

Assuming that the current levels in the wire are much higher than the critical current of the Josephson junctions $I_0$, i.e. $\Phi_0/(2\pi L_K)\gg I_0$, results in the following expression for the critical current of a SQUID with a ring included:
\begin{eqnarray}
&&I_c=2I_0\left|\cos\left(\pi f_{\mathrm SQUID}+\pi(f_{\mathrm ring}+n)\alpha\right)\right|\label{switch}\\
&&{\mathrm where}\quad\alpha=\frac{1+\frac{M_G}{L_K}}{\frac{s}{a}+\frac{L_G}{L_K}}
\end{eqnarray}
Here $f_{\mathrm SQUID}\Phi_0$ is the external flux in the SQUID loop and 
$f_{\mathrm ring}\Phi_0$ the external flux in the ring. $M_G$ is the mutual inductance between the ring and the SQUID loop.
The first part in the cosine argument gives the regular SQUID behavior. 
The second part is caused by the ring.
The critical current pattern is shifted by $n\pi\alpha$. Additionally the ring contributes 
$\alpha$ times the area of the ring to the oscillation period. Furthermore in derivation of equation \ref{switch} the effects of the self inductance of the SQUID loop are neglected. 

If the ring is small compared to the SQUID loop, the ratio between the mutual inductance $M_G$ and the self inductance $L_G$ of the ring 
 is equal to the ratio between the enclosed section $a$ and the total circumference $s$. 
Then the $\alpha$ factor is just equal to these ratios $\alpha=a/s=M_G/L_G$. 

In our experiment the ring is not small compared to the SQUID loop. The ratio $M_G/L_G$ for the $\pi$/2-SQUID is 0.20 and 
for the $\pi$-SQUID 0.35. To check the consistency of our experiment with theoretical expressions  
(\ref{switch}), one can calculate $\alpha$ from either
the period or the phase shift. For the $\pi$-SQUID we obtain $\alpha$=0.462 from the period
and $\alpha$=0.467 from the phase shift. For the $\pi$/2-SQUID we get $\alpha$=0.248 from the period
and $\alpha$=0.253 from the phase shift. 

We also cooled down with zero flux applied to the rings. All three devices showed regular SQUID 
behavior with a maximum switching current at zero field, i.e. zero phase shift. Equally consistent with predictions were the
patterns produced cooling down with minus one flux quantum and two flux quanta. 
We tested the stability of the frozen fluxoid by suddenly shorting and switching on the magnet 
supply, but we saw no change in the state.

The flux applied to the ring during
the cool down does not have to be exactly a flux quantum because $n$ is an integer. Deviating from 
the integer value will merely increase the chance of freezing a different number of fluxoids in the ring.

\begin{figure}[htb]
\begin{center}
\epsfig{figure=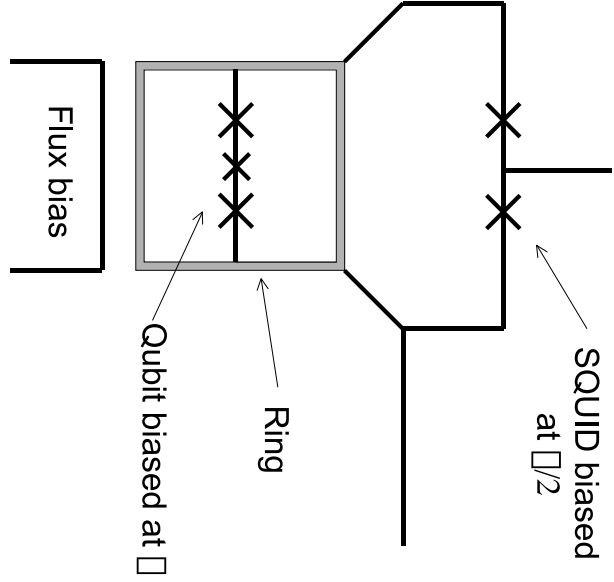, width=10 cm, clip=true}
\end{center}
\caption{Simplified flux qubit scheme. With a single ring the qubit is biased at $\pi$ and the read-out SQUID at $\pi/2$.}
\label{QubitScheme}
\end{figure}

Phase-bias rings can also be applied for flux qubits \cite{Mooij}, which need a shift of $\pi$ to reach the degeneracy point where superposition states occur. The ring bias is expected to have much lower flux noise than an external magnetic field. Fig.~\ref{QubitScheme} shows a simple version of a gradiometer qubit where both the qubit and the measuring SQUID are biased by the same ring.

\begin{acknowledgements}

We acknowledge useful discussions with P. Hadley, Y. Nakamura, C. J. P. M. Harmans, T. P. Orlando and A. Morpurgo and technical assistance from A. van der Enden and R. Schouten.
 This research was supported by Stichting voor Fundamenteel Onderzoek der Materie (FOM).

\end{acknowledgements}

\widetext

\begin{table}[h]
\caption{Results from a Fourier series fit of the measurements in Fig. (\ref{measurements}).}
\begin{tabular}{rrr}
Sample & Period & Phase \\
\hline
Reference SQUID & 118.2 $\mu$T & 0.1 $\mu$T\\
$\pi$/2-SQUID & 128.1 $\mu$T & 32.5 $\mu$T\\
$\pi$-SQUID & 120.5 $\mu$T & 56.4 $\mu$T
\end{tabular}
\label{fit}
\end{table}


\begin{references}
\bibitem{Tinkham}M. Tinkham, {\it Introduction to Superconductivity}, (McGraw-Hill, Newyork, 1996)

\bibitem{Ioffe} L. B. Ioffe, V. B. Geshkenbein, M. V. Fiegel'man, A. L. Fauchere, and
G. Blatter, Nature (London) {\bf 398}, 670 (1999).

\bibitem{Mooij}  J. E. Mooij, T. P. Orlando, L. Levitov, L. Tian, C. H. van der Wal, and
S. Lloyd, Sience {\bf 285}, 1036 (1999).

\bibitem{Blatter}  G. Blatter, V. B. Geshkenbein, L. B. Ioffe, Phys. Rev. B {\bf 63}, 174511 (2001) 

\bibitem{Beasley}  E. Terzioglu, M. R. Beasley, IEEE Trans. Appl. Supercond. {\bf 8}, 48 (1998). 

\bibitem{dwave} E. Il'ichev {\it et al}, Phys. Rev. Lett. {\bf 86}, 5369 (2001).

\bibitem{SFS} V. V. Ryazanov, V. A. Oboznov, A. Yu. Rusanov, A. V. Veretennikov, A. A. Golubov 
and J. Aarts, Phys. Rev. Lett. {\bf 86}, 2427 (2001). 

\bibitem{orlando} T. P. Orlando, K. A. Delin, {\it Foundations of Applied Superconductivity} (Addison-Wesley Publishing Company 1990)

\bibitem{mccumberhalperin} D. E. McCumber, B. I. Halperin, Phys. Rev. B. {\bf 1}, 1054 (1970).

\end{references}
\end{document}